\documentclass{ws-ijmpb}

\begin{document}

\markboth{Shiladitya Acharya and Krishnendu Mukherjee}
{HEAT TRANSPORT IN A THREE DIMENSIONAL SLAB GEOMETRY AND 
THE TEMPERATURE PROFILE OF 
INGEN-HAUSZ'S EXPERIMENT} 

\catchline{}{}{}{}{}

\title{HEAT TRANSPORT IN A THREE DIMENSIONAL SLAB GEOMETRY AND 
THE TEMPERATURE PROFILE OF INGEN-HAUSZ'S EXPERIMENT} 

\author{ Shiladitya Acharya}

\address{Department of Physics,\\ Bengal Engineering and 
Science University,\\ Shibpur, Howrah- 711 103, West Bengal, India\\
shiladitya.acharya@gmail.com}

\author{Krishnendu Mukherjee}

\address{Department of Physics,\\ Bengal Engineering and 
Science University,\\ Shibpur, Howrah- 711 103, West Bengal, India\\
kmukherjee@physics.becs.ac.in}

\maketitle

\begin{history}
\received{Day Month Year}
\revised{Day Month Year}
\end{history}


\begin{abstract}

We study the transport of heat in a three dimensional, 
harmonic crystal of
slab geometry whose boundaries and the intermediate surfaces 
are connected to stochastic, white noise heat baths at different 
temperatures. Heat baths at the intermediate surfaces 
are required to fix the initial state of the slab in respect of 
its surroundings. We allow the flow of energy fluxes between
the intermediate surfaces and the attached baths and 
impose conditions that relate the widths of Gaussian
noises of the intermediate baths. The radiated heat obeys 
Newton's law of cooling when
intermediate baths collectively constitute the environment 
surrounding the slab. We show that Fourier's law holds in the 
continuum limit.
We obtain an exponentially falling temperature profile from high 
to low temperature end of the slab and this very nature of the profile
was already confirmed by Ingen-Hausz's experiment. Temperature 
profile of similar nature is also obtained in the one dimensional version
of this model.

\end{abstract}


\section{Introduction}

We take a piece of solid bar and establish a steady 
temperature gradient between its two ends. 
The rest part of the bar remains exposed to the environment.
The transport of heat involves three processes such as,
conduction, radiation and absorption of heat. In 
the steady state absorption is absent and the 
conducted current density ${\bf J}({\bf x})$ obeys Fourier's law: 
\begin{equation}
 {\bf J}({\bf x})=-\kappa\nabla T({\bf x}),  
\label{Fourier}
\end{equation}  
where $\nabla T({\bf x})$ be the local temperature gradient 
and $\kappa$ is known as the thermal conductivity of the solid. 
In the steady state the temperature profile of the bar is 
exponentially falling from high to low temperature end and
this very nature was already confirmed by the experiment performed
by Ingen-Hausz\cite{Saha}.
Since the transport of heat is a non-equilibrium process,
it becomes a theoretical challenge to derive Fourier's law
and the exponentially falling temperature profile from the application
of non-equilibrium statistical mechanics to a basic model of solid.
It is observed that the heat transport in one dimension
exhibits anomalous\cite{Lepri1} behaviour. 
It means that the thermal conductivity shows a power law dependence 
$\kappa \sim N^{\alpha}$, where $N$ be the linear size of the system.
There are models which predict divergent 
($0<\alpha<1$) thermal conductivity
\cite{Rieder,Lepri1,Dhar1,Grassberger1,Narayan,Mai} in one dimension.
Numerical study indicates a logarithmic divergence\cite{Lippi} 
of thermal conductivity $\kappa\sim \ln{N}$ in two dimension. 
A power law behaviour\cite{Grassberger2} is also observed in 
such a system.
The validity of Fourier's law is confirmed numerically\cite{Dhar3} 
in one and two dimensional systems with pining and anharmonicity.
It has been shown numerically\cite{Chaudhuri} that 
a normal transport takes place in the  
disordered harmonic crystal in three dimension, subjected 
to an external pining potential.
The validity of this law is also established using 
a simulation study\cite{Dhar4} in three dimensional 
anharmonic crystal. This study provides a temperature
profile which is almost linear and very little concave
in the upward direction.

It is worthy to be mentioned of an interesting 
model of harmonic crystal
with 'self-consistent' stochastic heat baths\cite{Bonetto}.
A rigorous derivation shows that the Fourier's law holds in
dimension one of this model. However, for a $d$ dimensional crystal, when
$d$ is very large, the heat conductivity is shown to behave  
as $(l_d d)^{-1}$, where $l_d$ is the coupling to the intermediate 
reservoirs. Using 'self-consistent' heat baths one 
fixes the flow of energy fluxes between the 
intermediate sites and the attached baths to zero and 
these conditions give rise to a temperature profile
which falls linearly from high to low temperature 
end of the slab.
In this paper we use the same model   
in three dimension only, without the on-site binding potential
in the Hamiltonian. Instead of using 'self-consistent' baths,
we allow the heat to flow 
between the intermediate surfaces and the attached 
baths and impose conditions among the Gaussian 
widths of the noises of intermediate heat baths.
We show that under these conditions the radiated heat 
obeys Newton's law of cooling and the conducted heat satisfies
Fourier's law in the continuum limit. The temperature
profile exhibits an exponential fall from high to low
temperature end of the slab. A similar profile 
was obtained experimentally long before by Ingen Hausz\cite{Saha}.

We organize the paper in the following manner. We introduce the 
model and give the steady state solution of the Langevin's
equation in Section 2. We evaluate different correlators 
involving normal coordinates and normal velocities
in Section 3. We give the derivation of the exact temperature 
profile of the slab and discuss its experimental consequences 
in Section 4. This section also contains the derivation of
Newton's law of cooling and Fourier's law of heat conduction 
in the continuum limit. The results of one dimensional version
of this model are also discussed in this section. The discussion
and the conclusion are given in Section 5.

\section{Model}

We consider a cubic crystal in three dimension.
The form of the Hamiltonian  
\begin{eqnarray}
H = \sum_{\bf n} \frac{\dot x_{\bf n}^2}{2}+\sum_{\bf n,\hat{e}} \frac{1}{2} 
 (x_{\bf n} - x_{\bf n+\hat{e}})^2. 
\end{eqnarray} 
The displacement field $x_{\bf n}$ is 
defined on each lattice site
${\bf n}=(n_1,n_2,n_3)$ where $n_1=1,\cdots,N$, $n_2=1,\cdots,W_2$,
and $n_3=1,\cdots,W_3$.
Here ${\bf \hat{e}}$ denotes the unit vector in the three directions. 
We choose the value of mass attached to each lattice point and the 
harmonic spring constant as one.  
We have Langevin's type heat baths that are coupled to the 
surfaces at $n_1=1$ and $n_1=N$ and are maintained at fixed 
temperatures $T_1$ and $T_N$ ($T_1\,>T_N$) respectively.
Apart from these two heat baths, the surroundings
play a key role to fix the initial state of the slab.
Suppose, the heat baths are removed from the two end surfaces of the slab.
In this situation we expect that the temperature of the slab must 
be equal to the temperature of the surroundings in the stationary state 
limit. To achieve this expectation, in addition to two heat baths that 
are already fixed at the end surfaces of the slab, we require to couple 
$N-2$ more Langevin's type heat baths at the intermediate surfaces from 
$n_1=2$ to $n_1=N-1$. To deal with the same problem in one dimension
an equilibrium phase space distribution has been assumed to fix 
the initial state of the system\cite{Rieder}.
Now the equation of motion of a particle at the site ${\bf n}$ reads 
\begin{equation}
{\ddot{x}}_{\bf n}=-\sum_{{\bf \hat{e}}} (x_{\bf n}-x_{\bf n+\hat{e}})
-\gamma {\dot{x}}_{\bf n}\\
+ \eta_{\bf n}.
\label{eom}
\end{equation}
We have chosen the noises to be white and they are uncorrelated at
different sites. Assuming the process to be Gaussian, the noise 
correlation is specified by
\begin{equation}
\langle \eta_{\bf n}(t)
\eta_{\bf n^\prime}(t^\prime)\rangle
=\gamma z_{n_1} \delta(t-t^\prime)
\delta_{{\bf{n}},
{\bf{n}}^\prime}.
\label{noise1}
\end{equation}
According to ref.~\cite{Bonetto}, $z_{n_1}$ is chosen to be 
proportional to the temperature of the corresponding layer.
However, we determine them in our model using fluctuation dissipation 
theorem. We use the following periodic 
boundary conditions for the displacement field and the noise: 
\begin{eqnarray} 
x_{{\bf n}+(0,W_2,0)}(t) &=& x_{\bf n}(t) = x_{{\bf n}+(0,0,W_3)}(t)\nonumber\\
\eta_{{\bf n}+(0,W_2,0)}(t) &=& \eta_{\bf n}(t)
=\eta_{{\bf n}+(0,0,W_3)}(t)
\label{pbc}  
\end{eqnarray}
These boundary conditions lead to the following 
expansion for $x_{\bf n}(t)$ and $\eta_{\bf n}(t)$:  
\begin{eqnarray}
x_{\bf n}(t)&=&\frac{1}{\sqrt{W_2 W_3}}
\sum_{\bf{p}}  y_{n_1}({\bf{p}},t)
{\rm e}^{i{\bf{p}}.{\bf{n}}_\perp a},
\label{ftxn}\\
\eta_{\bf n}(t)&=&\frac{1}{\sqrt{W_2 W_3}}
\sum_{{\bf{p}}} f_{n_1}({\bf{p}},t)
{\rm e}^{i{\bf{p}}.{\bf{n}}_\perp a},
\label{ftetan}
\end{eqnarray} 
where $a$ be the lattice constant of the crystal,
${\bf{p}}=(p_2, p_3)$ and ${\bf{n}}_\perp=(n_2, n_3)$.
Upon substitution of Eq.~(\ref{ftxn}) and (\ref{ftetan}) into
Eq.~(\ref{eom}) we obtain
\begin{equation}
\ddot{y}_j=-\sum_{k=1}^N V_{jk}y_k-\gamma\dot{y}_j+f_j
\label{eom1}
\end{equation}
where the $N\times N$ matrix 
\begin{equation}
V=\left(\begin{array}{ccccc}
2\omega_0^2 & -1 & 0 & 0 & \ldots  \\ 
-1 & 2\omega_0^2 & -1 & 0 & \ldots  \\
0 & -1 & 2\omega_0^2 & -1 & \ddots \\
\vdots & \ddots & \ddots & \ddots & \ddots  \\ 
0 & \ldots & 0 & -1 & 2\omega_0^2
\end{array}\right)
\end{equation}
and 
\begin{equation} 
\omega_0^2({\bf{p}}) = 1+2\sin^2(\frac{p_2 a}{2})
+2\sin^2(\frac{p_3 a}{2}).
\label{w0}
\end{equation} 
We have assumed here
that $y_0({\bf{p}},t)=0=y_{N+1}({\bf{p}},t)$.
To solve Eq.~({\ref{eom1}}) we diagonalize the matrix $V$.
The solution of the $N$ order polynomial equation 
$\left|V-\alpha^2 I\right|=0$
gives the eigenvalues of $V$ as 
\begin{equation}
\alpha_k^2({\bf{p}})=2\omega_0^2({\bf{p}})+
2\cos\left(\frac{k\pi}{N+1}\right).
\label{eval} 
\end{equation}
The $j$-th component of the normalized eigenvector corresponding 
to the eigenvalue $\alpha_k^2$ is given by 
\begin{equation}
a^{(k)}_j=\sqrt{\frac{2}{N+1}} (-1)^{j+1} \sin\left(\frac{jk\pi}{N+1}\right).
\label{evec}
\end{equation} 
The diagonalizing matrix $A$ thus reads as
$A_{jk}=a^{(k)}_j$ such that $A^TA=I$ and 
$A^TVA=\alpha^2$, where $(\alpha^2)_{jk}=\alpha_j^2\,\delta_{jk}$.
We introduce a new set of coordinates $\xi_j$ as
\begin{equation}
y_j({\bf{p}},t)=\sum_{k=1}^N A_{jk}\xi_k({\bf{p}},t).
\label{yxi}
\end{equation}
The equation of motion of $\xi_j$ in matrix form can be written as 
\begin{equation}
\ddot{\xi}=-\alpha^2\xi-\gamma \dot{\xi} +\tilde{f},
\label{eomxi} 
\end{equation}
where $\tilde{f}=A^Tf$.
Since the matrix $V$ is diagonal with respect to $\xi_j$
($j=1,\cdots, N$), we henceforth refer them as 
normal coordinates. It is evident from this equation 
that each normal mode is acted on by 
dissipative force and noise offered by the heat baths. 
Moreover, one normal mode is coupled to the other through 
noise terms. The couplings among normal coordinates lead 
to the flow of energy from one mode to the other and 
thereby leading to a non-ballistic flow of heat in the slab. 
In the steady state limit ($t>>1/\gamma$) we are interested
in the particular solution of the equations of motion 
of $\xi$. We use the Fourier transform of
\begin{equation}
\xi_j(t)=\int_{-\infty}^\infty\frac{d\omega}{2\pi}\xi_j(\omega)
{\rm e}^{i\omega t}
~{\rm and}~
f_j(t)=\int_{-\infty}^\infty\frac{d\omega}{2\pi}f_j(\omega)
{\rm e}^{i\omega t}
\label{ftxif}
\end{equation}
in Eq.~(\ref{eomxi}) and obtain
\begin{equation}
\xi_j({\bf{p}},\omega)=
-\frac{\tilde{f}_j({\bf{p}},\omega)}
{\omega^2-\alpha_j^2({\bf{p}}) -i\gamma\omega}.
\label{xiomega}
\end{equation} 
Now upon substitution of Eq.~(\ref{xiomega}) into (\ref{ftxif}) 
we obtain
\begin{equation}
\xi_j({\bf{p}},t) = -\sum_{k=1}^N\int_{-\infty}^\infty\frac{d\omega}{2\pi}
\frac{{\rm e}^{i\omega t}}{\omega^2-\alpha_j^2({\bf{p}})-i\gamma\omega}
a^{(j)}_k\,f_k({\bf{p}},\omega).
\label{xiss}
\end{equation}
Now the use of Eq.~(\ref{ftetan}) and (\ref{ftxif}) 
in (\ref{noise1}) gives the noise correlation in frequency and 
wave-vector space as 
\begin{equation}
\langle f_j({\bf{p}},\omega)f_k({\bf{p}}^\prime,\omega^\prime)
\rangle = 2\pi\gamma z_j\delta_{j,k}\delta(\omega+\omega^\prime)
\delta_{{\bf{p}}+{\bf{p}}^\prime,0}.
\label{noise2}
\end{equation}

\section{Correlation functions}

For the purpose of calculating physically observable quantities
in this model, such as temperature profile and heat current 
density, we require to obtain the correlation functions involving normal
coordinates and velocities.
First we compute the correlation between normal coordinate and 
normal velocity using Eq.~(\ref{xiss}) and (\ref{noise2}). 
We perform a frequency integration using delta function in 
frequency space and obtain the correlation as
\begin{equation}
\langle \xi_{k_1}({\bf{p}},t)\dot\xi_{k_2}({\bf{p}}^\prime,t^\prime)\rangle 
= \gamma I_c(t-t^\prime) \delta_{{\bf{p}}+{\bf{p}}^{\prime},0}
\sum_{j=1}^N a_j^{(k_1)}a_j^{(k_2)} z_j
\label{correlation}
\end{equation}
where
\begin{eqnarray}
& &I_c(t-t^\prime)\nonumber\\
&=& -i\int_{-\infty}^\infty\frac{d\omega}{2\pi}
\frac{\omega {\rm e}^{i\omega(t-t^\prime)}}
{(\omega^2-\alpha_{k_1}^2-i\gamma\omega)
(\omega^2-\alpha_{k_2}^2+i\gamma\omega)}.
\end{eqnarray}  
Performing the integration over $\omega$ we obtain
\begin{equation}
I_c(t-t^\prime) = \frac{{\rm e}^{-\gamma|t-t^\prime|/2}}
{4\,\Delta_d(\beta_1,\beta_2)}[I_c^>(t-t^\prime)\theta(t-t^\prime)
+\,I_c^<(t-t^\prime)\theta(t^\prime-t)],
\label{ic}
\end{equation}
where
\begin{eqnarray}
\Delta_d(\beta_1,\beta_2) &=& (\cos\beta_1-\cos\beta_2)^2
+\,\gamma^2(2\omega_0^2({\bf{p}})
+\cos\beta_1+\cos\beta_2),\label{deld}\\
I_c^>(t-t^\prime) &=& 2(\cos\beta_1-\cos\beta_2)
\cos(\omega_{k_1}|t-t^\prime|)
+\,\frac{\gamma}{\omega_{k_1}}\{(4\omega_0^2+3\cos\beta_1\nonumber\\
& &+\cos\beta_2)
\sin(\omega_{k_1}|t-t^\prime|)\},\label{icg}\\
I_c^<(t-t^\prime) &=& 2(\cos\beta_1-\cos\beta_2)
\cos(\omega_{k_2}|t-t^\prime|)
-\,\frac{\gamma}{\omega_{k_2}}\{(4\omega_0^2+\cos\beta_1\nonumber\\
& &+3\cos\beta_2)
\sin(\omega_{k_2}|t-t^\prime|)\},\label{icl}\\
\beta_{1,2} &=& \pi k_{1,2}/(N+1),\label{beta12}\\
\omega_{k_{1,2}} &=& \sqrt{\alpha^2_{k_{1,2}}-\gamma^2/4}.
\label{omegak12}
\end{eqnarray}
It is clear that $I_c(t-t^\prime)\rightarrow 0$, when
$|t-t^\prime|\rightarrow\infty$. When $t=t^\prime$
\begin{equation}
I_c(0) = \frac{\cos\beta_1-\cos\beta_2}{2\,\Delta_d(\beta_1,\beta_2)}.
\label{ic0}
\end{equation}
We compute the equal time 
velocity-velocity correlation using Eq.~(\ref{xiss}) and (\ref{noise2}) 
and after performing the algebraic manipulation obtain the correlation as
\begin{eqnarray}
\langle\dot{\xi}_{k_1}({\bf{p}},t)
\dot{\xi}_{k_2}(-{\bf{p}},t)\rangle 
&=& \frac{\gamma^2}{N+1}
\frac{2\omega_0^2+\cos\beta_1+\cos\beta_2}
{\Delta_d(\beta_1,\beta_2)}\nonumber\\
& &\times\sum_{j=1}^N z_j\sin(j\beta_1)\sin(j\beta_2).
\label{vvcorrelation}
\end{eqnarray}

\section{Temperature profile and Fourier's law}

Consider a particle on the surface at $n_1$.
The mean square velocity
of the particle in the steady state limit reads 
\begin{eqnarray}
v_{m}^2(n_1) &=& \frac{1}{W_2W_3}
\sum_{{\bf{n}}_\perp}
\langle{\dot{x}}^2_{\bf{n}}\rangle\nonumber\\
&=&\frac{1}{W_2W_3}\sum_{{\bf{p}}}\sum_{k_1,k_2=1}^N
a_{n_1}^{(k_1)}a_{n_1}^{(k_2)}
\langle\dot{\xi}_{k_1}({\bf{p}},t)
\dot{\xi}_{k_2}(-{\bf{p}},t)\rangle,
\label{vav1}
\end{eqnarray}
where we have used Eq.~(\ref{ftxn}) and (\ref{yxi}) 
in the last step. Then we substitute Eq.~(\ref{evec}) and 
(\ref{vvcorrelation}) in the above equation   
and evaluate $p_2$ and $p_3$ sum in the continuum limit.
According to this limit, when $a\rightarrow 0$ and 
$W_{2,3}\rightarrow \infty$, $a\,W_{2,3}$ remain fixed 
and the discrete sums over $p_2$ and $p_3$ are converted into 
integrals:
\begin{equation}
\sum_{{\bf{p}}} \rightarrow \frac{a^2 W_2W_3}{(2 \pi)^2}
\int_{-\frac{\pi}{a}}^{-\frac{\pi}{a}}
\int_{-\frac{\pi}{a}}^{-\frac{\pi}{a}}
\,dp_2\, dp_3.
\end{equation}
Evaluation of the integrals\cite{Gradshteyn} over $p_2$ and $p_3$ 
gives
\begin{equation}
v_{m}^2(n_1)= 2 \sum_{j=1}^N C_{n_1,j} z_j
\label{vm2}
\end{equation}
where
\begin{eqnarray}
C_{n_1,j} &=& \frac{1}{(N+1)^2}\sum_{k_1,k_2=1}^N
\frac{\Lambda(\beta_1,\beta_2)}{\Delta(\beta_1,\beta_2)}\sin(n_1\beta_1)
\sin(n_1\beta_2)\sin(j\beta_1)\nonumber\\
& &\times\sin(j\beta_2),
\label{Cmatrix}\\
\Lambda(\beta_1,\beta_2) &=& \Delta(\beta_1,\beta_2)
- \{(\cos\beta_1-\cos\beta_2)^2\nonumber\\
& &\times F(1/2, 1/2, 1; (4\gamma^2/\Delta(\beta_1,\beta_2))^2)\},
\label{bigl}\\
\Delta(\beta_1,\beta_2) &=& (\cos\beta_1-\cos\beta_2)^2
+\gamma^2\,(6+\cos\beta_1+\cos\beta_2).
\label{delta}
\end{eqnarray}
The $N\times N$ matrix $C$ has the following properties:
\begin{equation}
C_{j,k}=C_{k,j}~~~{\rm and}~~~
C_{j,k}=C_{N+1-j,N+1-k}.
\label{propCjk}
\end{equation}
To thermalize the layers it is required that the average kinetic energy
of a particle on a layer in the steady state limit coincides  
with its energy at thermal equilibrium. This requirement
gives the following $N$ simultaneous equations of $z_k$:
\begin{equation}
\sum_{k=1}^NC_{j,k} z_k = \frac{1}{2} T_j,
\label{therm1}
\end{equation}
where we have chosen the Boltzmann constant $k_B=1$.
$T_j$ is the temperature of the $j$-th layer.
According to this equation, if $C_{j,k}=C_{j,j}\delta_{j,k}$, 
then only $z_j$ will be proportional to $T_j$\cite{Bonetto}.
This equation also indicates that the system in the steady state consists 
of $N$ different, thermally equilibriated sub-systems which are 
characterized by the temperatures $T_1, \cdots, T_N$. 
In Eq.~(\ref{therm1}), $T_1$ and $T_N$ are given. Apart from
these two, all $T_j$ for $2\le j\le N-1$ and $z_j$ for $1\le j\le N$
remain unknown. We require to know additional conditions 
which will facilitate us to obtain the temperature profile of the 
slab and the $z_j$s as a solution of these $N$ equations. 

Since $T_1>T_N$, there will be a rectilinear flow of heat 
from the left boundary surface at $n_1=1$ to the right 
boundary surface at $n_1=N$ of the slab. 
Heat current density $j_{\bf n}$ from the 
lattice site ${\bf n}$ to
${\bf n+\hat{e}_1}$, where ${\bf \hat{e}_1}=(1,0,0)$, 
is given by\cite{Lepri1}
\begin{eqnarray}
j_{\bf n}=\frac{1}{2}\langle (x_{\bf n}-x_{\bf n + \hat{e}_1})
(\dot{x}_{\bf n + \hat{e}_1}+\dot{x}_{\bf n})\rangle
\end{eqnarray}
The average heat current density per bond\cite{Dhar4} 
\begin{equation}
J = \frac{1}{W_2W_3 (N-1)}
\sum_{n_1=1}^{N-1}\sum_{n_2=1}^{W_2}\sum_{n_3=1}^{W_3} j_{\bf n}.
\end{equation}
We substitute Eq.~(\ref{ftxn}) and (\ref{yxi}) in $J$ and after 
performing the summations over $n_2$ and $n_3$ obtain the 
average heat current density per bond in the steady state limit as
\begin{eqnarray}
J&=&-\frac{1}{2W_2W_3 (N-1)}
\,\sum_{{\bf{p}}}\, \sum_{k_1,k_2=1}^N
\,\sum_{n_1=1}^{N-1}(a_{n_1+1}^{(k_1)} - a_{n_1}^{(k_1)})
(a_{n_1+1}^{(k_2)} + a_{n_1}^{(k_2)})\nonumber\\
& &\times\langle\xi_{k_1}({\bf{p}},t)\dot{\xi}_{k_2}(-{\bf{p}},t)
\rangle.
\label{hc0}
\end{eqnarray}
We now use Eq.~(\ref{evec}) and evaluate the sum over $n_1$. 
The result is given in the following: 
\begin{eqnarray}
\sum_{n_1=1}^{N-1}(a_{n_1+1}^{(k_1)} - a_{n_1}^{(k_1)})
(a_{n_1+1}^{(k_2)} + a_{n_1}^{(k_2)})
&=&\frac{2}{N+1}(1-(-1)^{k_1+k_2})\sin\beta_1\sin\beta_2\nonumber\\
& &\times\Big[\frac{1}{\cos\beta_2-\cos\beta_1}-1\Big].
\label{n1sum}
\end{eqnarray}
We use this result and also the results of 
Eq.~(\ref{correlation}) and (\ref{ic0}) 
in Eq.~(\ref{hc0}).
Then we evaluate the discrete sums over $p_2$ and $p_3$ in
the continuum limit and obtain
\begin{equation}
J=\frac{\gamma}{N-1} \sum_{j=1}^N z_j I_j(N,\gamma),
\label{hc1}
\end{equation}
where
\begin{eqnarray}
I_j(N,\gamma) &=& \frac{1}{(N+1)^2}\sum_{k_1,k_2=1}^N
\frac{(1-(-1)^{k_1+k_2})}{\Delta(\beta_1, \beta_2)}
\sin(j\beta_1)\sin(j\beta_2)\sin\beta_1\sin\beta_2\nonumber\\ 
& &\times F\left(\frac{1}{2},\frac{1}{2},1;
\,(4\gamma^2/\Delta(\beta_1,\beta_2))^2\right).
\label{ing}
\end{eqnarray}
Using the property that $I_{N+1-j}(N,\gamma)=- I_j(N,\gamma)$ and 
assuming that $N$ be an even number, Eq.~(\ref{hc1}) simplifies to
\begin{equation}
J = \frac{\gamma}{N-1}\sum_{j=1}^{N/2}(z_j - z_{N+1-j})\,
I_j(N,\gamma).
\label{hc2}
\end{equation}
It is known\cite{Bonetto} that there is also 
an energy flow between any intermediate 
surface and its attached heat bath. Since we are not 
using the self consistency condition, this intermediate 
energy fluxes will remain non-zero and   
affect the per bond average current density $J$. 
Consequently, $J$, as is evident from Eq.~(\ref{hc2}), 
is dependent on the variables, 
$\sqrt{\gamma z_2},\cdots, \sqrt{\gamma z_{N-1}}$,
the Gaussian widths of the noise functions 
of the intermediate heat baths. Assume that 
the widths are such that 
\begin{equation}
z_j=z_{N+1-j}~~{\rm for}~~2\le j\le N/2.
\label{zconstraints}
\end{equation}
The above conditions imposed on $z_j$s are physically interesting
firstly because of the fact that the use of them leads to a heat 
current density which obeys Fourier's law in the continuum limit.
Secondly, the use of them leads to a temperature profile whose 
nature is similar to what has been observed in Ingen-Hausz's
experiment. Thirdly, these conditions, unlike the self consistency
conditions used in Ref.~\refcite{Bonetto}, allow the radiation
to take place in the steady state limit of the transport process.
Now the use of these conditions in Eq.~(\ref{therm1}) gives the 
following relations:
\begin{eqnarray}
z_1-z_N &=& \frac{1}{2}\left(\frac{T_1-T_N}{C_{1,1}-C_{N,1}}\right),
\label{zrelation}\\
T_j-T_{N+1-j} &=& \frac{C_{j,1}-C_{N+1-j,1}}{C_{1,1}-C_{N,1}}(T_1-T_N)
\label{Trelation}
\end{eqnarray}
for $1\le j\le N/2$.
Moreover, the use of Eq.~(\ref{zconstraints}) in (\ref{therm1})
reduces $N$ equations into following $N/2$ simultaneous equations of 
the independent variables $z_j$ ($j=1, \cdots, N/2$):
\begin{equation}
\sum_{k=1}^{N/2}\bar{C}_{j,k}z_k=\frac{1}{2}\left(T_j
+\frac{C_{j,N}}{C_{1,1}-C_{N,1}}(T_1-T_N)\right),
\label{therm2}
\end{equation}
where $j=1, \cdots, N/2$ and
\begin{equation}
\bar{C}_{j,k}=C_{j,k}+C_{j,N+1-k}.
\end{equation}
\begin{figure}[bt]
\centerline{\psfig{file=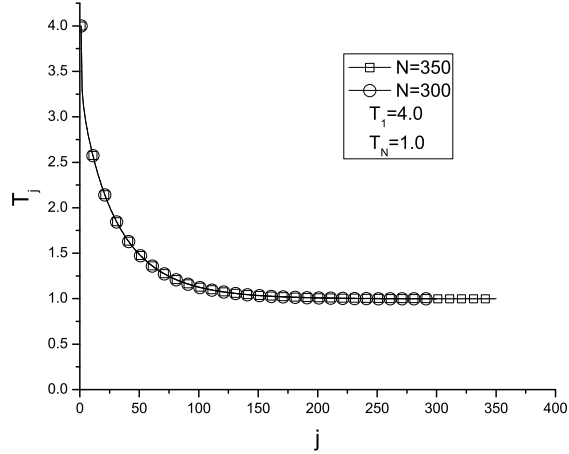,width=3.65in}}
\vspace*{8pt}
\caption{Temperature profile of the slab for $\gamma=0.01$.} 
\label{Figtp}
\end{figure}
To solve eq.~(\ref{therm2}) for $z_j$ ($1\le j\le N/2$), 
we require to know the temperature profile of the slab.
The temperatures of the boundary surfaces $T_1$ and $T_N$
are known and apart from these two, the relations given  
by eq.~(\ref{Trelation}) provide information
about the temperatures of the intermediate surfaces. 
The profile which is 
consistent with those relations must be 
$T_j=T_0+C_{j,1}(T_1-T_N)/(C_{1,1}-C_{N,1})$, 
where $T_0$ is a constant independent of $j$.
$T_0$ will be such that $T_j$ equals to $T_1$
and $T_N$ for $j$ equals to $1$ and $N$ respectively.
So it gives the temperature profile as
\begin{equation}
T_j = \left[1-\frac{C_{1,1}+C_{N,1}}{C_{1,1}-C_{N,1}}
\frac{T_1-T_N}{T_1+T_N}\right]
\frac{T_1+T_N}{2}
+\frac{C_{j,1}}{C_{1,1}-C_{N,1}}
(T_1-T_N),
\label{tprofile}
\end{equation}
where $1\le j\le N$. Moreover, the profile is in conformity with 
the zeroth law of thermodynamics which tells that $T_j=T_1$ for
all $j$ when $T_1=T_N$.
The temperature profile is plotted in 
Figure~(\ref{Figtp}). Since the profiles become size independent for 
$N > 300$, the open squares remain coincident with the open
circles till $j=300$. 
We use eq.~(\ref{tprofile})
in (\ref{therm2}) to solve $z_j$ for $j=1, \cdots, N/2$ and 
plugging these solutions into eq.~(\ref{zconstraints}) 
and (\ref{zrelation}) the rest of $z_j$ for $j$ lying in the range 
$N/2+1\le j\le N$ are obtained. The result of our numerical 
evaluations are given in the following table.
\begin{table}[!th]
\tbl{$z_j$ ($1\le j\le N$) for different $N$ and 
for $\gamma=0.01$.}
{\begin{tabular}{|c|c|c|}
\hline
$N$ & $z_1$ & $z_j~ {\rm for}~2\le j\le N $ \\
\hline
$300$ & $153.61$ & $1.99$ \\
\hline
$350$ & $153.45$ & $2.0$ \\
\hline
\end{tabular}}
\label{tabzj}
\end{table}
The Table~{\ref{tabzj}} indicates that $z_j$ are positive
definite and size independent for $N\ge 300$.

\begin{figure}[bt]
\centerline{\psfig{file=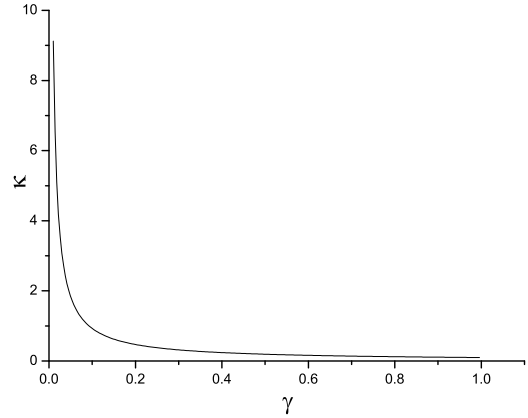,width=3.65in}}
\vspace*{8pt}
\caption{Plot of $\kappa$ as a function of $\gamma$.} 
\label{Figkg}
\end{figure}

We use the conditions of eq.~(\ref{zconstraints}) into
(\ref{hc2}) and then using eq.~(\ref{zrelation}) obtain
\begin{equation}
J=\frac{\gamma}{2(N-1)}
\left(\frac{T_1-T_N}{C_{1,1}-C_{N,1}}\right) I_1(N,\gamma).
\end{equation}
$I_1(N,\gamma)$ is zero if $k_1$ and $k_2$ simultaneously take even
integer values or odd integer values. 
Assuming that $N$ be an even
number and using the fact that the summoned of eq.~(\ref{ing}) is
symmetric in respect of the interchange of $\beta_1$ and $\beta_2$, 
we rewrite the double sum of $I_1(N,\gamma)$ as
\begin{equation}
I_1(N,\gamma) = \frac{4}{(N+1)^2}\sum_{j_1,j_2=1}^{N/2}
\frac{\sin^2\tilde{\beta}_1\sin^2\tilde{\beta}_2}
{\Delta(\tilde{\beta}_1,\tilde{\beta}_2)} 
 F\left(\frac{1}{2},\frac{1}{2},1;
\,(4\gamma^2/\Delta(\tilde{\beta}_1,\tilde{\beta}_2))^2\right),
\label{ing1}
\end{equation}
where $\tilde\beta_1=2\pi j_1/(N+1)$ 
and $\tilde\beta_2=\pi(2j_2-1)/(N+1)$. 
Again in the continuum limit we convert this double sum into integrals.  
Defining the integration variables
in this limit as $\theta_{1,2}=2\pi j_{1,2}/(N+1)$, we convert
the discrete sums into integrals:
\begin{equation}
\frac{2}{N+1}\sum_{j_{1,2}=1}^{N/2}\rightarrow
\frac{1}{\pi}\int_0^\pi d\theta_{1,2}.
\end{equation} 
$I_1(N,\gamma)$ thus takes the form
\begin{eqnarray}
g_1(\gamma) &=& \lim_{N\rightarrow\infty}I_1(N,\gamma)\nonumber\\
&=&\frac{1}{\pi^2}\int_0^\pi d\theta_1\int_0^\pi d\theta_2
\frac{\sin^2\theta_1\sin^2\theta_2}
{\Delta(\theta_1,\theta_2)} 
F\left(\frac{1}{2},\frac{1}{2},1;
\,(4\gamma^2/\Delta(\theta_1,\theta_2))^2\right).
\label{g1g}
\end{eqnarray}
Proceeding in the similar manner we obtain the following 
form of $C_{1,1}-C_{N,1}$ in the continuum limit as
\begin{eqnarray}
g_2(\gamma) &=& \lim_{N\rightarrow\infty}(C_{1,1}-C_{N,1})\nonumber\\
&=& \frac{1}{\pi^2}\int_0^\pi d\theta_1\int_0^\pi d\theta_2
\sin^2\theta_1\sin^2\theta_2\frac{\Lambda(\theta_1,\theta_2)}
{\Delta(\theta_1,\theta_2)}.
\label{g2g}
\end{eqnarray}
The steady state current density per bond thus reads in the
continuum limit as
\begin{equation}
J = \kappa\frac{(T_1-T_N)}{N-1},
\end{equation}
where the conductivity 
\begin{equation}
\kappa=\frac{\gamma}{2}\frac{g_1(\gamma)}{g_2(\gamma)}.
\label{kpgm}
\end{equation}
Here $\kappa$ is found to be finite and independent of the 
size of the system. So, as a consequence of the conditions  
given by eq.~(\ref{zconstraints}), Fourier's law holds 
in the continuum limit.
The thermal conductivity $\kappa$, as given by eq.~(\ref{kpgm}),
is plotted as a function of $\gamma$ in Figure~\ref{Figkg}.
Here $\gamma$ appears as a constant in the dissipative force 
term of the Langevin's equation. 
Physically it denotes a viscous force experienced by the
particles of Brownian like of the slab of 
crystal owing to collisions with the particles of fluid 
which seems to constitute the heat baths\cite{Chaikin,Reif}.
Moreover, owing to collisions, particles of the slab
will experience random forces or noises. 
Noise does a significant portion of 
its work in overcoming the viscous drag experienced by 
a particle of the slab and the rest will be flown as
heat energy to the neighbouring particles. Therefore,
the increase of $\gamma$ reduces the rate of flow of
heat from a particle to its neighbours and thereby reducing 
the thermal conductivity of the system. Hence, it justifies 
reasonably the nature of variation of $\kappa$ with $\gamma$ 
as shown in fig.~\ref{Figkg}.

According to eq.~(\ref{tprofile}) $T_j$ depends linearly on $C_{j,1}$.
We fit $C_{j,1}$ with the following exponentially falling 
function such a way that
\begin{equation}
C_{j,1}=a_0+a_1{\rm e}^{-b\, |j-1|},
\label{fitCj1}
\end{equation}
where $a_0$, $a_1$ and $b$ are supposedly $N$ dependent parameters. 
This parametrized form satisfies all the properties of $C_{j,k}$
given by eq.~(\ref{propCjk}). The numerical values of the 
parameters for different $N$ are given in the following table.
\begin{table}[!th]
\tbl{Fitting parameters involved with $C_{j,1}$
for different $N$ and for $\gamma=0.01$.}
{\begin{tabular}{|c|c|c|c|}
\hline
N & $a_0$ & $a_1$ & $b$ \\
\hline
$500$ & $1.48\times 10^{-5}$ & $4.9868\times 10^{-3}$ & $0.03220$\\
\hline
$600$ & $1.48\times 10^{-5}$ & $4.9872\times 10^{-3}$ & $0.03212$\\
\hline
$1000$ & $7.99\times 10^{-6}$ & $4.9881\times 10^{-3}$ & $0.03193$\\
\hline
$2000$ & $3.71\times 10^{-6}$ & $4.9887\times 10^{-3}$ & $0.03181$\\
\hline
$3000$ & $2.41\times 10^{-6}$ & $4.9889\times 10^{-3}$ & $0.03177$\\
\hline
\end{tabular}}
\label{Tparameters}
\end{table}
The substitution of eq.~(\ref{fitCj1}) into eq.~(\ref{tprofile})
gives the temperature profile of the slab in the 
thermodynamic limit as
\begin{equation}
T_j=T_N+(T_1-T_N){\rm e}^{-b|j-1|}.
\label{tprofile1}
\end{equation}
It is clear that $b$ is the only parameter which appears in the 
temperature profile of the slab in the thermodynamic limit 
and the Table~\ref{Tparameters} indicates that it 
approaches to a constant value in this limit.   
The expression for the profile given by eq.~(\ref{tprofile1}) 
is similar to what has been obtained in the 
Ingen-Hausz's experiment\cite{Saha}.
Now it is possible to figure out from the temperature 
profile of the slab the probable physical processes that may take 
place when heat transports through it. 
According to Refs.~\refcite{Saha}, the transport of
heat in the steady state limit involves the 
processes of conduction and radiation of heat. 
The situation, when the radiation is absent, the 
temperature falls linearly from left to right 
boundary surface. On the other hand,
if the radiation is allowed to take place, 
temperature falls exponentially and this very nature 
is observed in the Ingen-Hausz's experiment.  
Thus in this paper, allowing the energy fluxes to be flown 
between the intermediate surfaces and the corresponding
heat baths and the use of the conditions given by 
eq.~(\ref{zconstraints}), we are basically allowing the radiation 
to take place along with the conduction of heat 
in the steady state limit.
If these energy fluxes are not allowed to be flown
then a linear variation of temperature is obtained\cite{Bonetto}. 
It implies that the self consistency conditions do not allow 
the radiation to take place during the steady state transport 
of heat in the slab.

To understand the physical meaning of the conditions given in
eq.~(\ref{zconstraints}) it is important to compute the loss of 
heat from the slab. The flux of heat energy from particle at
${\bf{n}}$ to the attached heat bath reads\cite{Bonetto}
\begin{equation}
R_{{\bf{n}}}=\gamma\dot{x}^2_{{\bf{n}}}
-\dot{x}_{{\bf{n}}}\eta_{{\bf{n}}}.
\end{equation}
Now consider a layer at $n_1$. The average flux of heat from a
particle on this layer to the attached heat bath is given as
\begin{equation}
G(n_1)=\frac{1}{W_2W_3}\sum_{{\bf{n}}_\perp}
\langle R_{{\bf{n}}}\rangle.
\end{equation}
To evaluate this average we use eq.~(\ref{ftxn}), (\ref{ftetan}), 
(\ref{evec}) and (\ref{noise2}). Finally the use of 
eq.~(\ref{Cmatrix}) and (\ref{therm1}) leads to the average
heat flux in the thermodynamic limit as
\begin{equation}
G(n_1)=\gamma (T_{n_1}-z_{n_1}).
\label{heatflux}
\end{equation}  
Now the use of eq.~(\ref{zconstraints}) leads to the following 
relations
\begin{equation}
G(N+1-n_1)-G(n_1)=\gamma(T_{N+1-n_1}-T_{n_1})
~~{\rm for}~~2\le j\le N/2.
\label{Grelations}
\end{equation}
These relations are valid when
\begin{equation}
G(n_1)=\gamma(T_{n_1}-T_{env})
\label{Gn1}
\end{equation}
and 
\begin{equation}
z_{n_1}=T_{env}
\label{zn1}
\end{equation}
for $2\le n_1\le N-1$. Here $T_{env}$ is a constant and is assumed
to be equal to the average temperature of the environment.  
The choice of $z_{n_1}$ is actually motivated by our numerical 
evaluations in Table~\ref{tabzj} which suggests that $z_{n_1}$ remains
constant for $2\le n_1\le N-1$. The eq.~(\ref{Gn1}) indicates 
that the average flux of heat from any particle of a layer to 
its attached heat bath is proportional to the difference in temperature
between the layer and the environment and thus it obeys the Newton's
law of cooling, provided that intermediate heat baths collectively
constitute the environment surrounding the slab.   
Therefore, eq.~(\ref{zconstraints}) physically
implies that the radiated heat form the slab obeys Newton's law
of cooling. Since, the Fourier's law of heat conduction and
the Newton's law of cooling hold simultaneously in the transport process, we 
expect an exponentially falling nature of the temperature 
profile in the steady state limit\cite{Saha}.
This expectation is supported by the results of our numerical evaluation 
given by eq.~(\ref{fitCj1}) and the Table~\ref{Tparameters}. 
The obtained temperature 
profile, as given by eq.~(\ref{tprofile1}), indeed shows an exponentially
falling nature from left to right end of the slab.
    
We find quite a similar results in the one dimensional 
version of this model. The conditions similar to eq.~(\ref{zconstraints})
make the radiated heat to obey Newton's law of cooling and the 
conducted heat to satisfy Fourier's law in the thermodynamic limit.
Since, the sum over discrete momentum states in $2$ and 
$3$ directions are absent in one dimension, the conductivity
assumes the form
\begin{equation}
\kappa^{(1D)}=\frac{\gamma}{2}\frac{g_1^{(1D)}}{g_2^{(1D)}}
\end{equation} 
where 
\begin{eqnarray}
g_1^{(1D)} &=& \frac{1}{\pi^2}\int_0^\pi d\theta_1\int_0^\pi d\theta_2
\frac{\sin^2\theta_1\sin^2\theta_2}
{\Delta_d^{(1D)}(\theta_1,\theta_2)},\\
g_2^{(1D)} &=& \frac{\gamma^2}{\pi^2}\int_0^\pi d\theta_1\int_0^\pi
d\theta_2
\frac{\sin^2\theta_1\sin^2\theta_2}
{\Delta_d^{(1D)}(\theta_1,\theta_2)}
(2+\cos\theta_1+\cos\theta_2).
\end{eqnarray}
Here $\Delta_d^{(1D)}(\theta_1,\theta_2)=
(\cos\theta_1-\cos\theta_2)^2+\gamma^2(2+\cos\theta_1+\cos\theta_2)$.
The temperature profile exhibits an exponentially falling nature
as indicated by eq.~(\ref{tprofile1}) 
with a value of $b$ which approaches to $0.0171$ in the 
thermodynamic limit.

\section{Discussion and conclusion}

According to this model, the additional heat baths 
coupled at the intermediate
surfaces constitute the environment surrounding the slab geometry. 
The additional heat baths are essentially required to 
fix the initial state of the slab.
When $t >> 1/\gamma$, the slab will attain a unique
steady state which is a unison of $N$ different, thermally 
equilibriated sub-systems, characterized by the temperatures 
$T_1, \cdots, T_N$.
Instead of using self consistent reservoirs\cite{Bonetto},
we allow the heat to flow between the intermediate surfaces
and the corresponding attached heat baths and impose the conditions 
that the widths of the Gaussian noises of $j$-th and $N+1-j$-th 
baths are same for $j=1, \cdots, N/2$, where $N$ being assumed to be even.    
Those conditions lead to interesting physically admissible consequences.
Firstly, the radiated heat obeys Newton's law of cooling
in the thermodynamic limit.
Secondly, Fourier's law is satisfied in the 
continuum limit. 
Consequently, we  
obtain a temperature profile, which unlike showing a linear variation 
of Refs.~\refcite{Bonetto} falls exponentially from left to
right end of the slab and this very nature of the profile
is also in conformity with that of Ingen-Hausz's experiment.


\begin{thebibliography}{99}

\bibitem{Rieder}Z. Rieder, J. L. Lebowitz and E. Lieb,
 \it{J. Math. Phys.} {\bf{8}}, 1073-1078 (1967). 
\bibitem{Lepri1}S. Lepri, R. Livi and A. Politi, 
 \it{Phys. Rep.} {\bf{377}}, 1-80 (2003). 
\bibitem{Dhar1}A. Dhar,
 \it{Phys. Rev. Lett.} {\bf{86}}, 3554-3557 (2001). 
\bibitem{Grassberger1}P. Grassberger, W. Nadler, and L. Yang,
 \it{Phys. Rev. Lett.} {\bf{89}}, 180601-1-4 (2002).
\bibitem{Narayan}O. Narayan and  S. Ramaswamy,
\it{Phys. Rev. Lett.} {\bf{89}}, 200601-1-4 (2002).
\bibitem{Mai}T. Mai, A. Dhar and O. Narayan, 
\it{Phys. Rev. Lett.} {\bf{98}}, 184301-1-4 (2007).
\bibitem{Lippi}A. Lippi and R. Livi, 
\it{J. Stat. Phys.} {\bf{100}}, 1147-1172 (2000).
\bibitem{Grassberger2}P. Grassberger and L. Yang, 
 (2002), preprint, http://arxiv.org/cond-mat/0204247. 
\bibitem{Dhar3}A. Dhar, 
\it{Adv. Phys.} {\bf{57}}, 457-537 (2008).
\bibitem{Chaudhuri}A. Chaudhuri, A. Kundu, D. Roy, A. Dhar,
J. L. Lebowitz and H. Spohn,
\it{Phys. Rev. B} {\bf{81}}, 064301-1-17 (2010).
\bibitem{Dhar4}K. Saito and A. Dhar,
\it{Phys. Rev. Lett.} {\bf{104}}, 040601-1-4 (2010).
\bibitem{Bonetto}F. Bonetto, J. L. Lebowitz and J. Lukkarinen,  
\it{J. Stat. Phys.} {\bf{116}}, 783-813 (2004).
\bibitem{Gradshteyn}I. S. Gradshteyn and I. M. Ryzhik, 
\it{Tables of Integrals, Series and Products},
6th ed. (Academic Press, New Delhi, 2001).
\bibitem{Saha}M. N. Saha and B. N. Srivastava,
\it{A Treatise on Heat}, pp 462-465, 5th ed. 
(The Indian Press, Allahabad, 1969). 
\bibitem{Chaikin}P. M. Chaikin and T. C. Lubensky,
\it{Principles of condensed matter physics}, 
(Cambridge University Press, New Delhi, 2009). 
\bibitem{Reif}F. Reif, \it{Fundamentals of Statistical and
Thermal Physics}, (McGraw-Hill, Singapore, 1985).
 
\end{thebibliography}
\end{document}